\documentstyle[prc,aps,preprint]{revtex}
\tightenlines
\begin{document}
\draft
\title{Effect of the source charge on charged-boson interferometry}
\author{T.~D.~Shoppa$^1$, S.~E.~Koonin$^2$, 
and R.~Seki$^{2,3}$}
\address{$^1$TRIUMF, Vancouver, BC V6T 2A3, Canada\\
$^2$W.~K.~Kellogg Radiation Laboratory, 106-38\\
California Institute of Technology, Pasadena, CA 91125\\
$^3$Department of Physics and Astronomy\\
California State University, Northridge, CA 91330}
\date{\today}
\maketitle

\begin{abstract}

We investigate quantal perturbations 
of the interferometric correlations of charged 
bosons by the Coulomb field of an instantaneous, charged source. 
The source charge increases the apparent source size by weakening 
the correlation at non-zero relative momenta. 
The effect is strongest for pairs with a small total momentum and 
is stronger for kaons than for pions of the same momenta.  
The experimental data currently available are well described by this 
effect without invoking Pratt's exploding source model.  A simple expression 
is proposed to account for the effect.
\end{abstract}

\bigbreak

Intensity interferometry of charged particles from heavy-ion
collisions is an important probe of 
the fireball created in such collisions.  The source parameters 
(lifetime, size, etc.) 
are important in determining the energy density of the source, from  
which the equation of state can be extracted\cite{PrattCZ}.
Unfortunately, the charged particles suffer Coulomb and strong interactions 
with each other and also with their source, complicating the interpretation 
of the interference patterns observed. 
The approximate Gamow correction for the Coulomb interaction between detected 
pions works well in the limit of the small source size\cite{GyulassyCC}, 
and improved corrections have been proposed for extended sources 
\cite{PrattCoulomb,Bowler}.
The effect of strong interactions between the two emitted particles on their 
correlations has been examined for
two emitted pions or kaons\cite{PrattCZ},
for a pair of the pion and kaon\cite{DK}, 
and for the case of two emitted protons\cite{Koonin}.

These calculations ignore the interactions of the emitted particles 
with the source.  The effect of strong interactions with a source was
investigated for pion-source absorptive optical potentials 
by Chu {\it et al.}\cite{Ming} using eikonal wave functions.  
But because of its long range and the large charge of the colliding ions, 
the Coulomb influence of the source is also expected to be significant, 
but study has been limited to a qualitative 
discussion~\cite{Baym} and to classical considerations\cite{CL}.  
In this letter, we report the first quantum-mechanical investigation 
of the effect of
the Coulomb interaction between the emitted particles and their source; 
specifically for two-boson interferometry of instantaneous sources.
We find that the Coulomb interaction tends to increase the apparent source size, 
in contradicting with the previous investigations\cite{Baym}\cite{CL}.  We also 
find that this apparent ``magnification'' can be evident in the experimental 
correlations binned by the total momentum of the pair.

The correlation between a pair of emitted particles is expressed 
in terms of their asymptotic momenta, ${\bf p}_1$ and ${\bf p}_2$. 
The correlation function depends on various factors: 
the symmetries between the emitted particles, 
the forces among the emitted particles and the source,
and the size and lifetime of the source.
Here, we treat the source instantaneous and massive, and examine 
the case of boson pairs;  ${\bf p}_1$ and ${\bf p}_2$ are thus 
measured in the rest frame of the source, which does not recoil.
The correlation function is expressed as
\begin{equation}
C({\bf p}_1,{\bf p}_2) = {
{
\int d {\bf r_1} d {\bf r_2}
g({\bf r}_1) g({\bf r}_2)
|\Psi^{(-)}_{{\bf p}_1,{\bf p}_2}({\bf r}_1,{\bf r}_2)|^2
}
\over
{
\int d {\bf r_1} g({\bf r}_1)|\phi^{(-)}_{{\bf p}_1}({\bf r}_1)|^2
\int d {\bf r_2} g({\bf r}_2)|\phi^{(-)}_{{\bf p}_2}({\bf r}_2)|^2
}
                         }.
\label{KooninPratt}
\label{eqg}
\end{equation}
where the bosons are assumed to be emitted without correlation 
from the source with a momentum-independent semiclassical probability 
$g({\bf r})$.  
$\phi^{(-)}_{\bf p}({\bf r})$ is the one-body scattering wave-function of 
a boson with asymptotic momentum $\bf p$ and incoming scattered wave, 
due to the interaction with the source.
$\Psi^{(-)}_{{\bf p}_1,{\bf p}_2}({\bf r}_1,{\bf r}_2)$ is the
two-body scattering wave-function of the bosons with the asymptotic momenta 
${\bf p}_1$ and ${\bf p}_2$.  

To clarify the effect of the source-pair interaction, 
we neglect the Coulomb and strong interactions
between the emitted pair,
assuming that these can be treated by the methods in q.~\cite{GyulassyCC},
\cite{PrattCoulomb}, and \cite{Bowler}.
$\Psi^{(-)}_{{\bf p}_1 , {\bf p}_2} ({\bf r}_1 , {\bf r}_2)$
is then expressed as a symmetrized product of $\phi^{(-)}_{\bf p}({\bf r})$'s,
\begin{equation}
\Psi^{(-)}_{{\bf p}_1 , {\bf p}_2} ({\bf r}_1 , {\bf r}_2)
 = 2^{-1/2} [ \phi^{(-)}_{{\bf p}_1} ( {\bf r}_1 ) 
              \phi^{(-)}_{{\bf p}_2} ( {\bf r}_2 )
             +\phi^{(-)}_{{\bf p}_1} ( {\bf r}_2 ) 
              \phi^{(-)}_{{\bf p}_2} ( {\bf r}_1 ) ].
\label{GeneralWF}
\end{equation}
Note that Eqs.~(\ref{KooninPratt}) and (\ref{GeneralWF}) give    
$C({\bf p}_1,{\bf p}_2) = 2$ at relative momentum,
${\bf q} \equiv {\bf p}_1 - {\bf p}_2 = 0$, 
as is the case of no final state interaction with source.  
For further clarity, we take the source charge distribution to be Gaussian
and define the apparent source size, $R_{\rm app}$, as
\begin{equation}
{ {\rm{d}^2 C(q)} \over {\rm{d} q^2} }|_{q=0} = - {R_{\rm app}}^2.
\label{defineapparentsize}
\end{equation}
In general, $R_{\rm app}$ will also depend upon the total momentum 
of the emitted pair, ${\bf P}= {\bf p}_1 + {\bf p}_2$.  Note that 
$R_{\rm app}$ as defined becomes the exact source size, $R_0$,
when there is no source-pair interaction.  

With no further approximation, we solve the Schr\"{o}dinger equation 
numerically for $\phi^{(-)}_{\bf p}({\bf r})$, 
using the partial-wave decomposition method
with the appropriate kinematics and boundary conditions.
Figure 1 shows $R_{\rm app}$ thus calculated, illustrating
the effect of a source charge
on the correlations of both positive and negative pion pairs.
We see that {\it in both cases, the apparent size is larger than the exact
size for small $P = |{\bf P}|$.}
Also, as suggested in Ref.~\cite{Baym}, the apparent size can become 
either larger or smaller than the exact size, depending on the charge of
the pair being either negative or positive.
It does, however, only by a very slight amount and at high $P$.
We make two further comments on Fig. 1: 1) At a fixed $P$, $C(q)$ is  
of a Gaussian form as a function of $q$. 
2) When $R_{\rm app}$ is integrated over ${\bf P}$ with the thermal 
spectrum of pions, 
the resultant effective $R_{\rm app}(T)$ is greater than $R_0$ 
(except for ${\pi}^+$ pairs at high temperatures, $kT >$ 400 MeV.) 

A simplified description using the eikonal approximation
reveals the essential physics of why the apparent size is larger than 
the true size.  It provides a useful 
expression for the apparent size as a function of $P$. 
In the eikonal approximation, 
\begin{equation}
\phi^{(-)}_{\bf p}({\bf r}) = \exp (i {\bf p} \cdot {\bf r} +
i \chi({\bf p},{\bf r}) ), 
\end{equation}
with
\begin{equation}
\chi({\bf p},{\bf r}) = -{ {1} \over {v} }
\int_\infty^0 V({\bf r} + z \hat{\bf p} ) dz,
\label{eikonalintegral}
\end{equation}
where $V(\bf r)$ is the Coulomb potential between the source and the emitted 
boson of the mass $m$ and the velocity $v$. 
Equation~(\ref{GeneralWF}) yields  
\begin{equation}
|\Psi( {\bf p}_1,{\bf p}_2, {\bf r}_1 , {\bf r}_2)|^2 
  = 1 + \cos({\bf q} \cdot ({\bf r}_1 - {\bf r}_2)
      + \chi({\bf p}_1,{\bf r}_1) + \chi({\bf p}_2,{\bf r}_2)
      - \chi({\bf p}_2,{\bf r}_1) - \chi({\bf p}_1,{\bf r}_2) ).
\label{eikonalwf2}
\end{equation}
Note that $\chi({\bf p},{\bf r})$ of Eq.~(\ref{eikonalintegral}) diverges
logarithmically\cite{Glauber}, but Eq.~(\ref{eikonalwf2}) converges 
at least as fast as $|{\bf r}_1 - {\bf r}_2| / |{\bf r}_1 + {\bf r}_2|$
because of the cancellation between $\chi$'s.
Numerically we find, for example, that the contribution beyond 15 fm
is negligible for the source size less than 8 fm.

Let us now consider the special case where the bosons 
emitted with parallel momenta.   The apparent size of a spherically symmetric 
charged source can then be expressed as 
\begin{equation}
R_{\rm app}^2(P) = R_{0}^2 + { {m} \over {P^2} } T +
                            {\left({ {m} \over {P^2} } \right)}^2 U^2. 
\label{leikonalsize}
\end{equation}
Here, $T$ and $U$ depend only on the source distribution and potential:
\begin{eqnarray}
T & = &  
   \int  d {\bf r}_1 d^3 {\bf r}_2 g({\bf r}_1) g({\bf r}_2)
   ({\bf r}_{12} \cdot \hat{\bf z})
   \left[ I({\bf r}_1) - I({\bf r}_2) \right] \\
\label{tineikonal}
U^2 & = & 4
   \int  d {\bf r}_1 d^3 {\bf r}_2 g({\bf r}_1) g({\bf r}_2)
   {\left[ I({\bf r}_1) - I({\bf r}_2) \right]}^2, 
\label{uineikonal}
\end{eqnarray}
where $I({\bf r}) = - v \chi({\bf p},{\bf r})$ 
is a function of the emission point, ${\bf r}$.
A close examination of Eqs.~(\ref{leikonalsize}) -- (\ref{uineikonal})
shows~\cite{timthesis} that the crucial
quantity is essentially the ratio of the Coulomb and kinetic energies,
\begin{equation}
     \zeta = { {Z e^2} \over {R_0}}/{ {P^2} \over {2m} },
\end{equation} 
which becomes large for large $Z$ and low $P$.

The $T$ term in Eq.~(\ref{leikonalsize}) is a linear functional of 
the Coulomb potential between
the source and the emitted pair, and thus either increases or decreases
$R_{\rm app}$, depending on the emitted pair either negatively or
positively charged. 
The $U$ term is, however, quadratic in the Coulomb potential and
{\it increases} $R_{\rm app}$ in all cases.  
For the cases considered below, $|\zeta|$ is significant 
when $P$ is smaller than a few hundred MeV/c, or even less.
The $U$ term can then contribute appreciably to $R_{\rm app}$, and
actually become dominant for a relatively wide range of small $P$.
$R_{\rm app}$ rises very rapidly ($\sim P^{-4}$) for the low $P$.
The contribution like the $T$ term is included in the discussion 
of Ref.\cite{Baym}, but the contribution like the $U$ term is not. 

Equations~(\ref{leikonalsize}) -- (\ref{uineikonal}) shows that
for a given $P$, the effect is stronger for heavier particles,
for example, kaons in comparison to pions.

The parameters $T$ and $U$ are proportional to the source charge in the 
eikonal approximation, and depend to a small extent on the source size
and geometry. The exact expression of $R_{\rm app}$ for arbitrary
${\bf P}$ and ${\bf q}$ is more complicated than Eq.~(\ref{leikonalsize}). 
Numerical calculations show, however, that these observations obtained 
for the special case of the eikonal approximation still hold, and that 
Eq.~(\ref{leikonalsize}) yields a quite good representation of the $P$ 
dependence of $R_{\rm app}$ if we treat $T$ and $U$ as adjustable 
parameters.  In this case, $T$ and $U$ follow semi-quantitatively these 
observations, including rising with increasing source charge roughly 
linearly.  In detail, $U$ representing the exact results (i.e. using exact wave 
functions) is smaller than those obtained with the eikonal wave function, 
especially for a large source charge.  
This is because $U$ is determined dominantly in the small $P$ region, 
where the eikonal wave function tends to overestimate the phase shifts.
However, $T$ is insensitive to small $P$ and is calculated reasonably 
reliably by the eikonal approximation.

More importantly, Fig.~1 suggests that the effect of the source-pair 
interaction can be seen clearly when the data are binned according to the 
total momentum of the pair, $P$.  Though yet relatively crude, such data are 
available.  
Figure~2 show comparison of our exact calculation with data 
at 650 MeV/nucleon~\cite{Bock} for 
the positive boson pairs, $\pi^+  \pi^+$ and $K^+ K^+$.  
For the sake of clarity, we show  
the calculation in Figs. 2 and 3 after smoothing out by means of Eq.~(7) 
the curves obtained by numerical integration as those in Fig.~1. 
In the upper panel of Fig.~2, 
the calculated effect of a $Z=158$ source charge 
is compared with $R_{\rm app}$ extracted from the $\pi^+ \pi^+$ 
data of Au $+$ Au central collisions, 
binned according to $P$ of the pair.  
We see that the calculation with the (true) size of $R_0 = 1.0$ fm 
reproduce the data well.  In the lower panel of Fig.~2, 
similar calculated effect of a $Z=82$ source charge 
with $R_0 = 2.8$ fm is compared with the Nb $+$ Nb collision data.  
The agreement between our calculation and the data is also good.

Figure 3 shows similar comparison for $\pi^- \pi^-$ pairs at a higher energy 
of 1.8 GeV/nucleon.  
In the upper panel, the calculation for a $Z=100$ source with   
$R_0 = 2.8$ fm is compared with the Ar $+$ Pb collision data\cite{BeavisPb}; 
and in the lower panel, the calculation for a $Z=36$ source with $R_0 = 4.2$ fm 
is compared with the Ar $+$ KCl collision data\cite{Beavis}.   The agreement 
appears to be reasonable.

In these calculations, $R_0$ is taken to be the asymptotic value 
of $R_{\rm app}$ at large $P$.
This is the only adjust parameter in the case of our exact calculation. 
Note that the rise of the apparent size at small $P$ is determined 
by the charge of the source, the mass and charge of the outgoing 
particles, and weakly depends on $R_0$.

The Ar $+$ KCl data of Fig.~3 has previously
been fit with an exploding source model \cite{PrattExplode}.
This model, which neglects the charge of the source but does
takes into account a radial expansion of the source, also predicts
an apparent source size dependence that will show larger sources
for small total momentum pairs.
The dotted line in the lower panel of Fig.~3 shows the exploding source 
fit of \cite{PrattExplode}.
This fit has two free parameters, a source size of 7.75 fm and a
expansion rate $T/\gamma v$ of 100 MeV.
In contrast, the predicted effect of the source charge we show here
only has one free parameter, the actual source size $R_0$; we take the
charge of the source from the total charge of the colliding
ions.  Even though the
general trend of the exploding source model is similar to
the results for a charged source, the exploding source
model shows a gentle behavior
at small total momentum $P$ and still has a significant slope at
large $P$.  The effect of the source charge is very strong at low total
momentum, and is very small at large total momentum.  The different
total momentum dependence of these two effects would allow them
to be disentangled, if experimental data finely binned by total
pair momentum becomes available.

This work is supported by the National Science Foundation grant at
Caltech (PHY-9412818 and PHY-9420470),
and by the U. S. Department of Energy grant
at CSUN (DE-FG03-87ER40347).

\vfill
\pagebreak

\vfill

\pagebreak

\begin{figure}
%\begin{center}
%\epsfig{file=FIGECURVESA.PS,width=6.0in}
%\end{center}
\caption{
The apparent source size $R_{app}$ as a function of the total momentum 
$P$ of $\pi^+$ pairs (upper panel) and of $\pi^-$ pairs (lower panel) emitted
from a source of size $R_0=$ 4 fm.  $R_{app}$ is computed by a numerical 
integration of the Schr\"{o}dinger equation.  The dotted lines are 
the $R_{app}$ 
for the source charge $Z=15$, and the dashed lines for $Z=100$.
The solid straight lines show $R_{app} = R_0 =$ 4 fm for $Z=0$.
}
%\label{FigEcurvesA}
\end{figure}

\begin{figure}
%\begin{center}
%\epsfig{file=FIGBOCK2.PS,width=6.0in}
%\end{center}
\caption{
The apparent source sizes from $\pi^+\pi^+$ interferometry of 
Au$+$Au (upper panel) and Nb$+$Nb (lower panel) collisions 
at 650 MeV/nucleon, binned according to the total momentum $P$. 
The horizontal error bars of the data\protect\cite{Bock}  
represent the binning according to total momentum.
The solid line represents the calculation obtained by numerically solving 
the Schr\"{o}dinger equation (and by being smoothed by means of Eq.~(7))  
for $Z=158$ (upper panel) and for $Z=82$ (lower panel). 
The dashed lines are the similar calculation for $K^+K^+$ pairs.
}
%\label{Bock2}
\end{figure}

\begin{figure}
%\begin{center}
%\epsfig{file=FIGBEAVIS2.PS,width=6.0in}
%\end{center}
\caption{
The apparent source sizes from $\pi^-\pi^-$ interferometry of 
Ar$+$Pb (upper panel) and Ar$+$KCl (lower panel) collisions 
at 1.5 GeV/nucleon, binned according to the total momentum $P$. 
Data of \protect\cite{BeavisPb} and \protect\cite{Beavis} 
in the upper and lower panels, respectively, show bin sizes. 
The calculation shown with the solid lines is by the numerical 
integration as in Fig.~2.  The exploding source fit 
of \protect\cite{PrattExplode} is shown by the dotted line 
in the lower panel.
}
\end{figure}


\begin{thebibliography}{10}

\bibitem{PrattCZ}
S. Pratt, T. Cs{\"{o}}rg{\H{o}}, and J. Zim{\'{a}}nyi, Phys. Rev. {\bf C42},
  2646  (1990).

\bibitem{GyulassyCC}
M. Gyulassy, S.~K. Kauffmann, and L.~W. Wilson, Phys. Rev. {\bf C20},  2267
  (1979).

\bibitem{PrattCoulomb}
S. Pratt, Phys. Rev. {\bf D72},  72  (1986).

\bibitem{Bowler}
M.~G. Bowler, Phys. Lett. {\bf B270},  69  (1991).

\bibitem{DK}
D.~J. Dean and S.~E. Koonin, Phys. Lett. {bf B305}, 5 (1993).

\bibitem{Koonin}
S.~E. Koonin, Phys. Lett. {\bf B70},  43  (1977).

\bibitem{Ming}
M.~C. Chu, S. Gardner, T. Matsui, and R. Seki, Phys. Rev {\bf C50},  3097
  (1994).

\bibitem{Baym}
G.~Baym and P. Braun-Munzinger, Nucl. Phys. {\bf A610}, 286c (1996).

\bibitem{CL}
D. Hardtke and T. J. Humanic, Phys. Rev. {\bf C57}, 3314 (1998);
H.~W. Barz, J. P. Bondorf, J. J. Gaardhoje, and H. Heiselberg, 
nucl-th/9711064.

\bibitem{Glauber}
R.~J. Glauber,  in {\em Lectures in Theoretical Physics}, edited by W.~E.
  Brittin and D.~G. Dunham (Interscience, New York, 1959), Vol.~1, p.\ 315.

\bibitem{timthesis}
T. D.~Shoppa, Ph.D thesis, Caltech (1996), unpublished.

\bibitem{Bock}
R. Bock {\it et~al.}, Mod. Phys. Lett {\bf A3},  1745  (1988).

\bibitem{Beavis}
D. Beavis {\it et~al.}, Phys. Rev. {\bf C27},  910  (1983).

\bibitem{BeavisPb}
D. Beavis {\it et~al.}, Phys. Rev. {\bf C34},  757  (1986).

\bibitem{PrattExplode}
S. Pratt, Phys. Rev. Lett. {\bf 53},  1219  (1984).

\end{thebibliography}
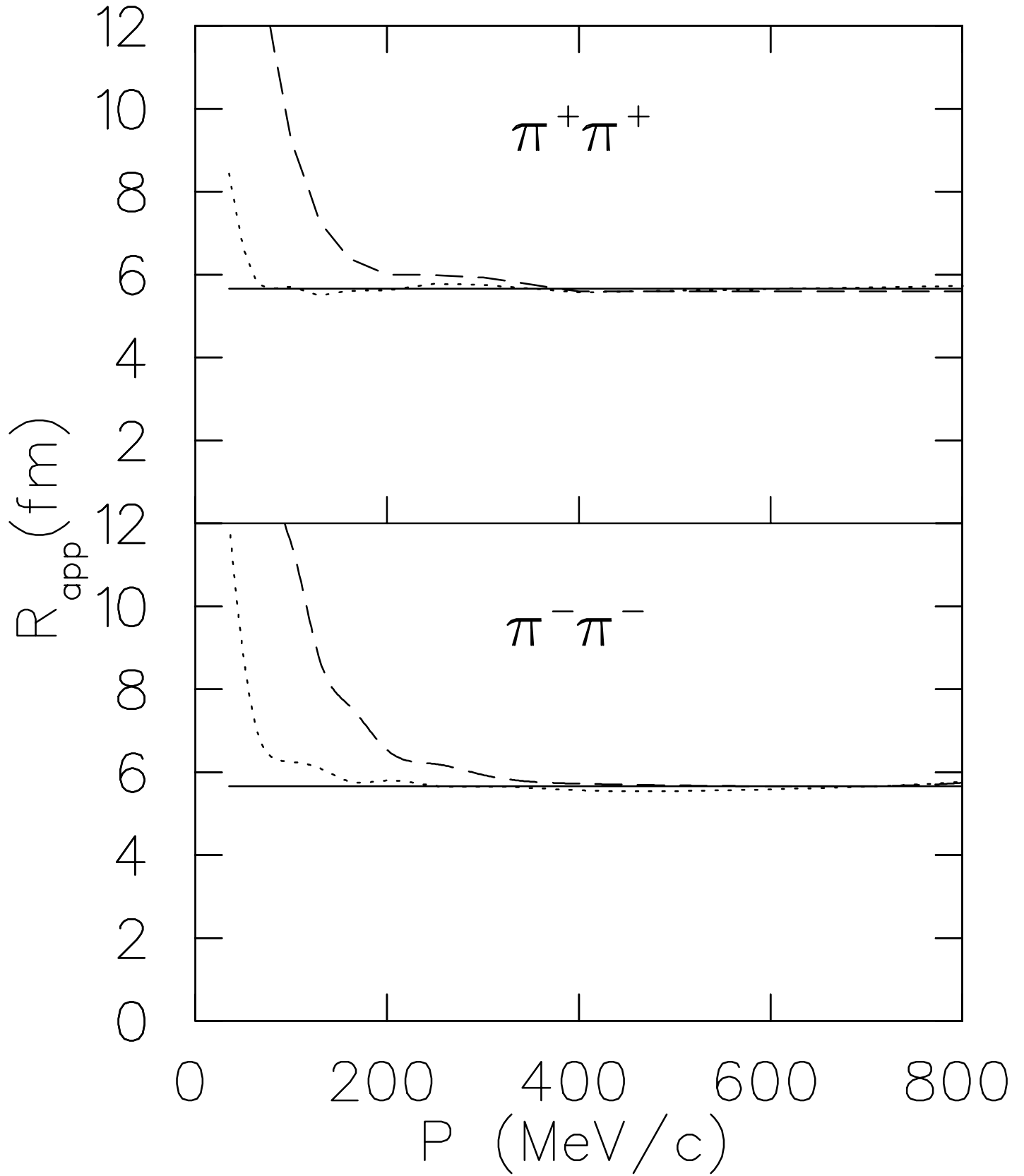
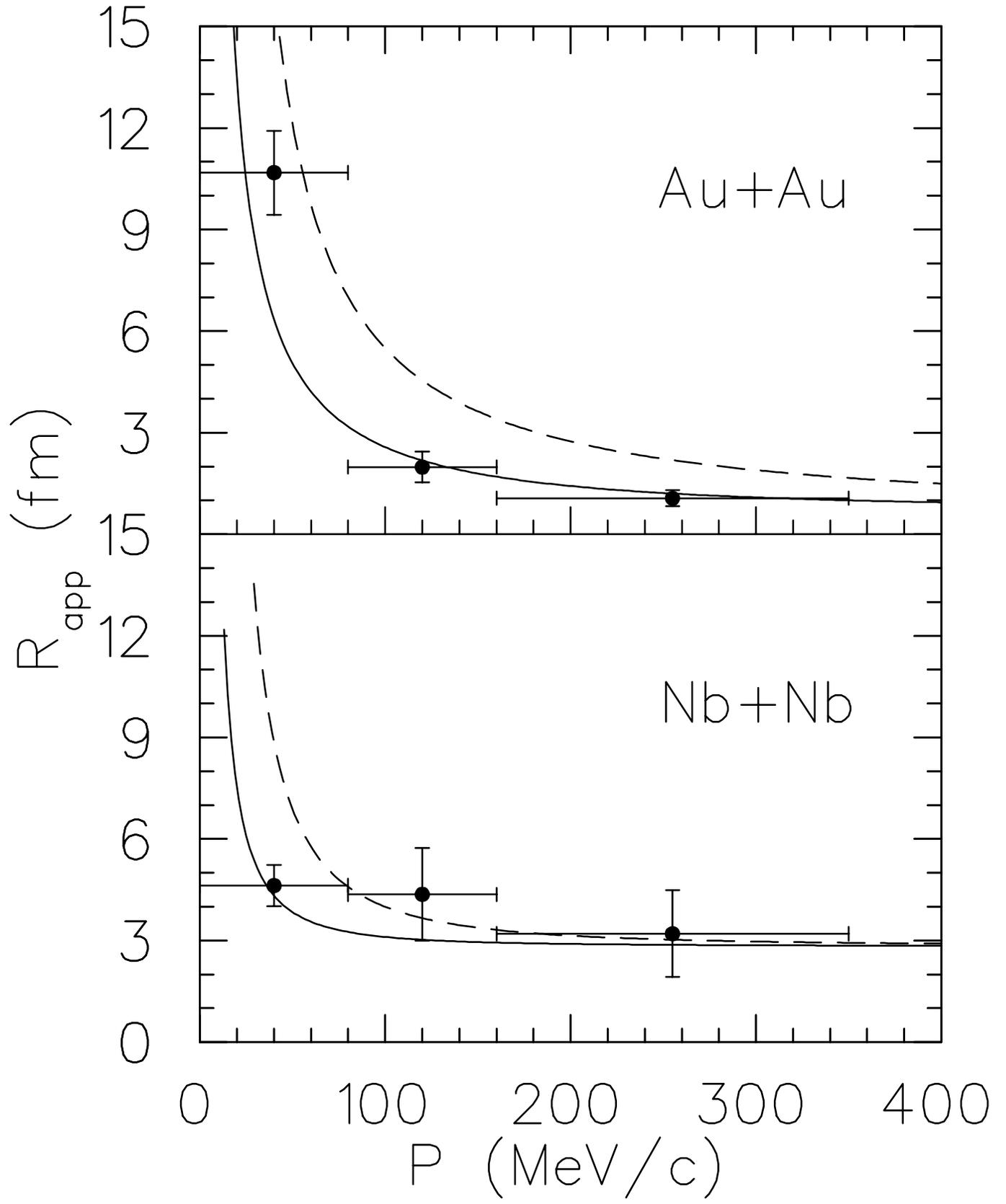
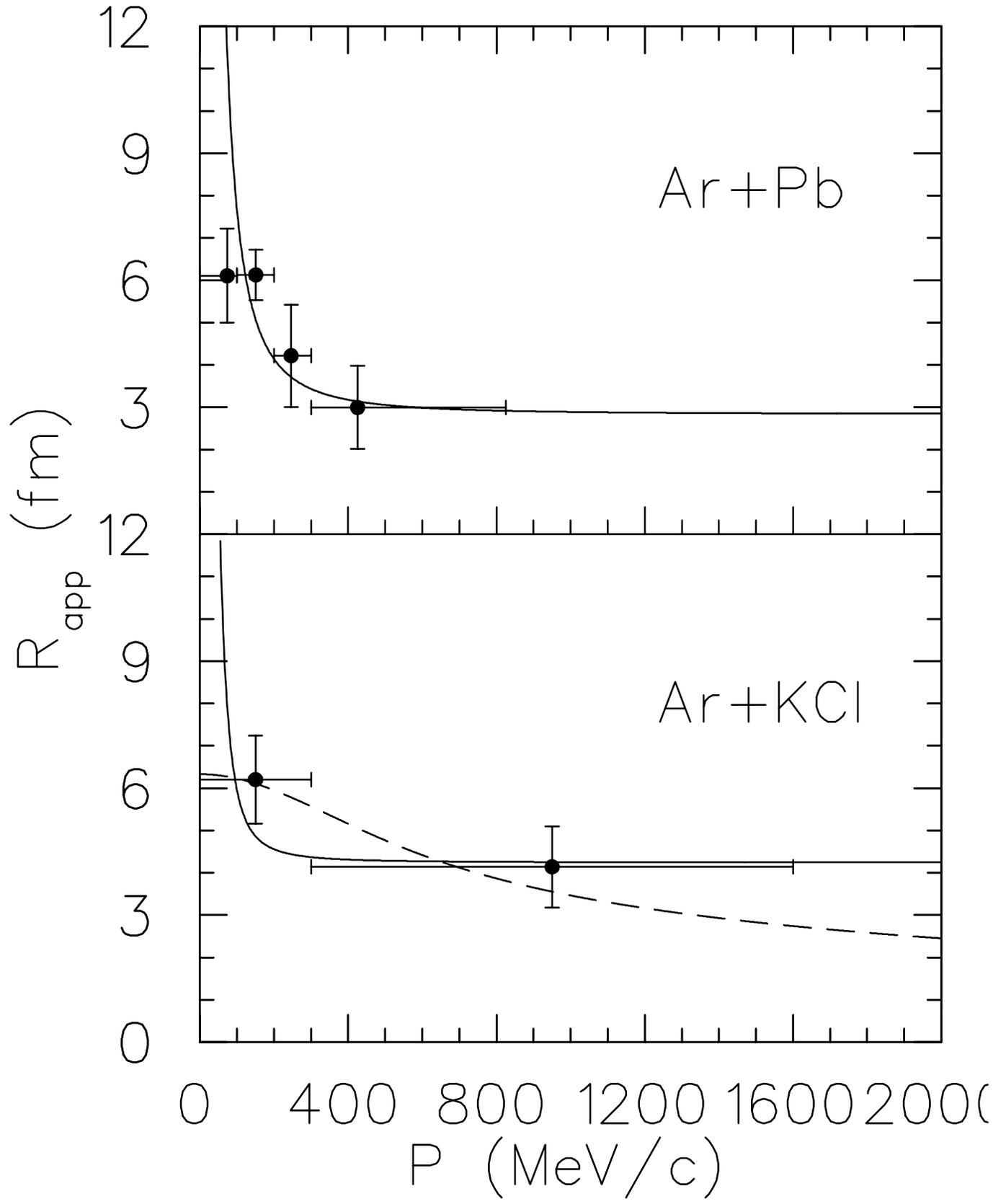
\end{document}